\documentclass[namedreferences]{solarphysics}

\usepackage[hyperref,optionalrh,showbiblabels]{spr-sola-addons} 
\usepackage{graphicx}        
\usepackage{color}           
\usepackage{breakurl}        

\chardef\us=`\_

\begin{document}

\begin{article}
\begin{opening}

\title{Interaction of CME/ICME with HSS solar wind from coronal holes: case study}

\author[addressref=aff1,corref,email={yermol@iki.rssi.ru}]{\inits{Yu.I.}\fnm{Yu.I.}~\lnm{Yermolaev}}
\author[addressref=aff1,email={irina-priem@mail.ru}]{\inits{I.G.}\fnm{I.G.}~\lnm{Lodkina}}
\author[addressref=aff2,email={jshugai@srd.sinp.msu.ru}]{\inits{Yu.S.}\fnm{Yu.S.}~\lnm{Shugay}}
\author[addressref=aff3,email={slem@sci.lebedev.ru}]{\inits{V.A.}\fnm{V.A.}~\lnm{Slemzin}}
\author[addressref={aff1,aff2},email={veselov@dec1.sinp.msu.ru}]{\inits{I.S.}\fnm{I.S.}~\lnm{Veselovsky}}
\author[addressref={aff3},email={rodkindg@gmail.com}]{\inits{D.G.}\fnm{D.G.}~\lnm{Rodkin}}
\author[addressref={aff1},email={deceased}]{\inits{N.S.}\fnm{N.S. }~\lnm{Nikolaeva}}
\author[addressref={aff1},email={nlbor@mail.ru}]{\inits{N.L.}\fnm{N.L.}~\lnm{Borodkova}}
\author[addressref={aff1},email={michaely2@yandex.ru}]{\inits{M.Yu.}\fnm{M.Yu.}~\lnm{Yermolaev}}


\address[id=aff1]{Space Research Institute (IKI), Russian Academy of Sciences, Moscow, Russia}
\address[id=aff2]{Skobeltsyn Institute of Nuclear Physics, Lomonosov Moscow State University, Moscow, Russia}
\address[id=aff3]{P.N. Lebedev Physical Institute, Russian Academy of Sciences, Moscow, Russia}

\runningauthor{Yu.Yermolaev et al.}
\runningtitle{Interaction of ICME with SW from CH}

\begin{abstract}

In this paper we  look for traces of interaction of CME/ICME with high-speed stream (HSS) of solar wind from coronal hole (CH) for two serial Carrington rotations (CRs) during December 2011 -- January 2012. We  analyze two cases. (1) In CR 2118 instead of  predicted arrival of HSS the signatures of ICME with modified back side and compressed slow stream were observed. (2)  In CR 2119 the predicted HSS was observed in time, and after arrival of preceding ICME. Our analysis shows that difference of these cases can be explained by different interactions between HSS and CME, which in the first case result in significant deflection of HSS of solar wind from the Sun-Earth line. Thus, the CH passage through the central meridian with simultaneous occurrence of CMEs in the nearby located sources can result in interaction between HSS and CME streams, deflection of the HSS stream aside of the ecliptic and changes in parameters of the solar wind.

\end{abstract}
\end{opening}

\section{Introduction}
     \label{S-Introduction}

As well known, the interaction of non-stationary large-scale phenomena of solar corona and solar wind as with surrounding plasma, and as with each other leads to noticeable change of a trajectory of their movement in the corona and interplanetary space. For example, the non-stationary solar phenomena like Coronal Mass Ejection (CME) can interact with such large-scale structures as streamers, coronal holes (CHs), other CMEs and solar wind (see, for instance, papers by \cite{gopals04,mishra15,mostl15} and references therein) . Authors of several works \citep{gopals09,mohamed12} show that coronal holes can change the CME trajectory, in particular, polar coronal holes deflect the CMEs to the equator. However, the back impact of CMEs on high-speed streams (HSS) from coronal holes has not been investigated in detail.

So far, the models for prediction the solar wind parameters in the near-Earth space and behind consider HSS and transient CME streams as independent features. The Wang-Sheeley-Arge model (WSA, \cite{wang90,arge00}) or EUV-imaging based models (e.g. \cite{vrsnak07,shugay11}) use location and geometry of coronal holes for simulation of quasi-stationary streams of slow solar wind and HSS. The kinematic models of CME propagation like Advanced Drag model \citep{vrsnak13}, consider propagation of CMEs in the heliosphere as a stream of a fast transient stream on a slowly moving homogeneous background wind with acceleration or deceleration depending on the relation between their velocities. The MHD-based WSA-ENLIL prediction model \citep{odstrcil99} is a combination of the WSA model in the corona producing initial conditions at the boundary between the corona and heliosphere (at 21 -- 30 $R_{sun}$), with the ideal MHD model, which describes propagation of quasi-stationary solar wind in the heliosphere. A modification of this model -- WSA-ENLIL cone model \citep{mays15} provides an estimation of CME arrival time using additional information about initial parameters of CME from coronagraphic observations. None of these models describes effects of probable interaction between magnetic structures of HSS and CME in the cases when they propagate with different velocities and meet in the heliosphere before arriving to Earth.

In this work we present an example of two consecutive Carrington rotations (CRs) of the Sun during December 2011 -- January 2012 (CRs 2118 and 2119) when CMEs happened before approach of a coronal hole to the Sun-Earth line: in one case (23 -- 24 January 2012) the high-speed stream was observed near Earth orbit, and in other case (29 -- 30 December 2011) it was not observed. In accordance with prediction on the basis of coronal hole observations, HSSs should be measured near the Earth in both cases.  We suggest that the CME can deflect the HSS relative to the Sun-Earth line and the HSS absence near the Earth is a result of CME-CH interaction.  We discuss the interplanetary data to confirm this hypothesis.

\section{Methods}
\label{Methods}
\subsection{Determination of the CME/ICME and HSS solar wind sources}
\label{Determination of the CME/ICME and HSS SW sources}

To understand the nature of probable interaction between HSS from CHs and transient CME streams during the process of their expansion in the corona and propagation in the heliosphere, one needs to consider the sources of these streams, which can be identified by specific signatures. HSS streams originate in CHs, which can be identified as wide regions of open magnetic field lines or seen in the EUV wavelength range (typically, in the 193 A line) as wide dark areas (Figure 1, the upper panel). To calculate the probable velocity of quasi-stationary HSS streams as well as to predict their arrival times to the Earth, we used the EUV-imaging method as described in \citet{shugay11,slemzin15} and references therein.

CMEs originate due to spontaneous solar activity with such signatures as X-ray flares and flux robe ejections in active regions (ARs), filament/prominence eruptions, dimmings etc., which can be used for localization of their solar sources. For this purpose, we used the data about CMEs directed to the Earth and their solar sources from the Solar Terrestrial Relations Observatory STEREO--A and B\footnote{http://helio.gmu.edu/seeds/secchi.php} \citep{howard08}, Large Angle and Spectrotrometric Coronagraph LASCO\footnote{http://cdaw.gsfc.nasa.gov} \citep{brueckner95}, Solar Dynamic Observatory SDO/AIA telescope\footnote{http://sdo.gsfc.nasa.gov} \citep{lemen12}. CMEs are identified as directed to the Earth if: (1)their position angles in the coronagraphic images of STEREO--A and B lie in the proper half--plane (in our case at the eastern limb for Cor2A and at the western limb for Cor2B); (2)their angular half--widths are large than the angles between the direction of propagation and the ecliptic plane. Parameters of solar flares we took from the Solar Demon database\footnote{http://solardemon.oma.be/science/flares.php} \citep{kraaikamp15}.

For calculations of the CME arrival times we used two heliospheriс CME-propagation models: the Advanced Drag Model (ADM, \cite{vrsnak13}), and the WSA-Enlil cone model \citep{odstrcil99, pizzo11}, presented in the internet\footnote{http://helioweather.net}. To discover the effect of interaction between HSS and CME streams, we compared the measured in-situ velocities with those simulated by the CH-based model and the WSA-ENLIL model (Figure 1, the bottom panel).

\subsection{Determination of the solar wind phenomena} 
  \label{Determination of the SW phenomena}

In our work we analyze the following solar wind phenomena: the high-speed stream (HSS), the interplanetary coronal mass ejection (ICME) and compression region Sheath before ICME.  To identify the solar wind phenomena we use the standard parameter criteria: (1) HSS has  the bulk speed V $>$ 450 km $s^{-1}$, (2) ICME has low ratio of thermal to magnetic pressures (beta-parameter) and high and smoothly rotating magnetic field, and (3) Sheath has high  temperature, density and beta-parameters. Numerical criteria and procedure of phenomenon identification are described in details in paper \citep{yermolaev09}, the results of identifications of solar wind phenomena are presented in the internet\footnote{ftp://ftp.iki.rssi.ru/pub/omni/catalog/}.

Recently we analyzed the interplanetary magnetic field (IMF) and plasma parameters of the OMNI dataset\footnote{http://omniweb.gsfc.nasa.gov} for period 1976 -- 2000 and calculated the average temporal profiles of several parameters for 8 usual sequences of phenomena: (1) SW/ CIR/ SW, (2) SW/ IS/ CIR/ SW, (3) SW/ Ejecta/ SW, (4) SW/ Sheath/ Ejecta/ SW, (5) SW/ IS/ Sheath/ Ejecta/ SW, (6) SW/ MC/ SW, (7) SW/ Sheath/ MC/ SW, (8) SW/ IS/ Sheath/ MC/ SW (where SW is undisturbed solar wind and CIR is corotating interaction region) \citep{yermolaev15} using the method of double superposed epoch analysis for large numbers of events \citep{yermolaev10}. These data can be used as templates to compare the measured parameters of selected phenomena with average temporal profiles of corresponding parameters.  We change the time between data points a such way that durations of selected events are equal to durations of corresponding average events in paper \citet{yermolaev15} and compare selected events with the sequence of phenomena SW/IS/Sheath/Ejecta/SW.

\section{Results and Discussion} 
  \label{Results and Discussion}

In this section, we consistently describe observations of the Sun and measurement of parameters of the interplanetary space near Earth.

\subsection{Interaction of CME with HSS on the basis of solar observations} 
  \label{Interaction of CME with HSS on the basis of solar observations}

We consider the cases of probable interaction of HSS from the recurrent CH and several CMEs appeared during the periods 21 December 2011 -- 04 January 2012 and 19 -- 31 January 2012 (two sequential Carrington rotations CR 2118 and 2119). The data on the sources of these transient and quasi-stationary solar wind streams and the times of their appearance at the near-Earth orbits were listed in Tables 1 and 2. On the upper panels of Figure 1 there are the synoptic maps constructed from the SDO/AIA EUV images in the 193 A channel with the encircled locations of the CME sources. On the bottom panels there are the predicted values of the solar wind velocity calculated by the HSS-predicting model based on identification of the CH areas from the EUV-imaging \citep{shugay11,slemzin15} and by the WSA-ENLIL cone model in comparison with the solar wind measurements on the ACE spacecraft. It is worth to pay attention on a significant difference between simulations by the WSA-ENLIL and the HSS-predicting models for the periods under study related with propagation of CMEs, since only the first model uses information about CMEs.

In CR 2118 HSS from the southern CH expected in the period from 30 December 2011 to 04 January 2012 was not registered by ACE. Instead, the signatures of ICME have been detected which can be associated with several CMEs originated from the active regions marked in Figure 1 (upper panel).  During the period 21 -- 28 December 2011, several CMEs with flares occurred in all three ARs, part of them being propagated in the direction of ACE. The CMEs originated from ARs 1384-1386 on 25 -- 27 December 2011 and directed to ACE, as it follows from the STEREO observations, would reach the Earth on 28 -- 31 December. At the same time, the HSS from CH has been expected (see Tables 1, 2 and Figure 1, the lower panel). Instead, in CR 2119 the fast component of solar wind arrived two days earlier than the HSS-predicting model expected it. The simulation with the WSA-ENLIL cone model suggests that this early appearance can be associated with arrival of ICME associated with the CME on 23 January 2012 from AR1402.

\begin{figure} 
\centerline{
\includegraphics[width=1.0\textwidth,clip=]{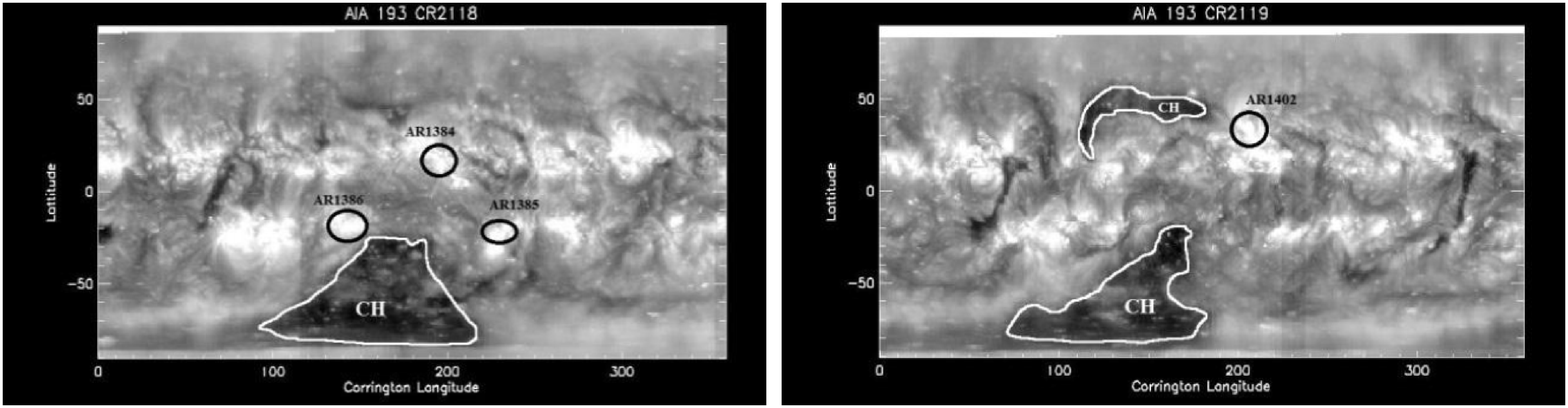}
}
\centerline{
\includegraphics[width=1.0\textwidth,clip=]{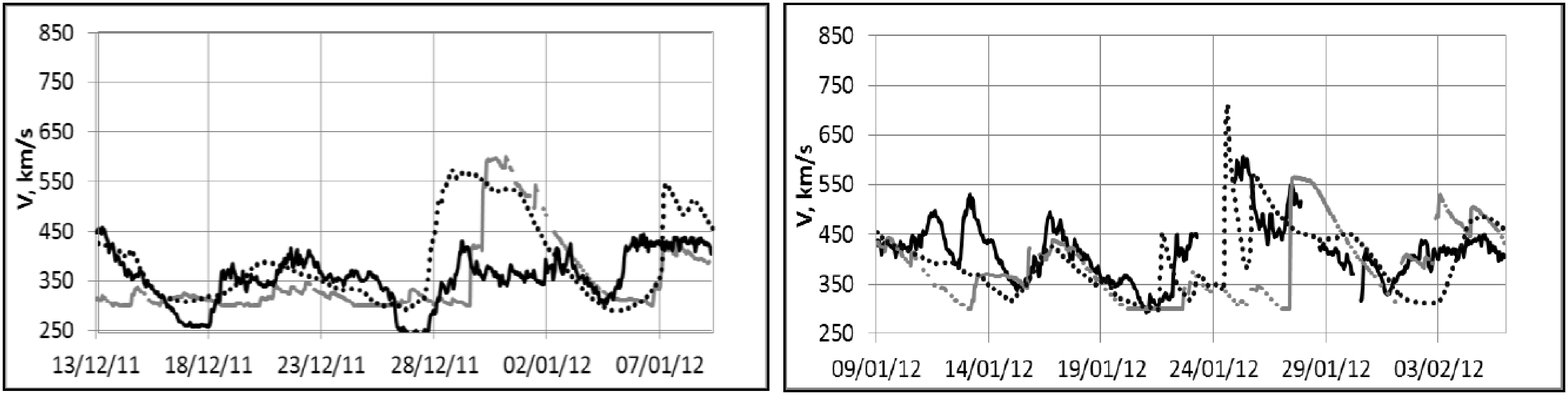}
}
\caption{Top panel: AIA synoptic maps at wavelength 193 {\AA} (http://spaceweather.gmu.edu/projects/synop/AIASM.html), with marked CHs (HSS SW sources) and ARs (CME sources). The bottom panel shows the SW speed measured by ACE (the black solid curve), calculated for HSS from the CH areas ( the gray curve) and calculated by WSA-ENLIL model (the dotted curve). Data presented for CR 2118 (left) and CR 2119 (right)
}
\label{fig1}
\end{figure}

\begin{table}  
\caption{CH and predicted HSS SW}
\label{CH and predicted HSS SW}
\begin{tabular}{ccccccc}     
\hline                   
& CH crossing the & Predicted HSS & Predicted $V_{max}$  \\
N & central meridian & arrival time & \\
& (at latitude $<$ 30$^{\circ}$) & & \\
\hline
 & start / end & start / end & Time & V km $s^{-1}$ \\
\hline
1 & 2011 Dec. 26 08:00 /  & 2011 Dec. 30 07:00 / & 2011 Dec. 31 & 597 \\
 & 2011 Dec. 29 13:00 & 2012 Jan. 4 16:00 & 09:00 & \\
\hline
2 & 2012 Jan. 23 17:00 / & 2012 Jan. 27 05:00 / & 2012 Jan. 27 & 633 \\
 & 2012 Jan. 25 17:00 & 2012 Jan. 31 12:00 & 11:00 & \\
 \hline

\end{tabular}
\end{table}

\begin{table}   
\caption{Flares, CMEs and predicted ICME arrival time.}
\label{Flares, CMEs and predicted ICME arrival time}
\begin{tabular}{cccc}     
\hline
Flare & CME & CME & ICME near the Earth  \\
SolarDemon & Lasco (CDAW) & Stereo & DBM / Enlil \\
Time & Time & Onset Time & \\
Class and AR & & STA (A) and STB (B)& \\
\hline
\bf{CR 2118} & & &   \\
\hline
- & 2011 Dec. 25 01:25 & 2011 Dec. 25 01:24 & 2011 Dec. 29 04:09 / \\
 & & & 2011 Dec. 28 20:00 \\
\hline
2011 Dec. 25 & 2011 Dec. 25 18:50 & (A)2011 Dec. 25 18:24 & 2011 Dec. 31 12:37 / \\
18:14 -- 19:16 & & (B)2011 Dec. 25 19:54 & 2011 Dec. 28 20:00 \\
M3 AR1385 & & & \\
\hline
2011 Dec. 26 & 2011 Dec. 26 11:48 & (A)2011 Dec. 26 11:24 & 2011 Dec. 30 02:40 / \\
11:32 -- 13:24 & & (B)2011 Dec. 26 12:24 & 2011 Dec. 29 01:00 \\
C7 AR1384 & & & \\
\hline
2011 Dec. 27 & - & (A)2011 Dec. 27 05:24 & 2011 Dec. 30 15:17 / \\
04:14 -- 05:14 & & (B)2011 Dec. 27 04:54 & - \\
M1 AR1386 & & & \\
\hline
\bf{CR 2119} & & &   \\
\hline
2012 Jan. 19 & 2012 Jan. 19 14:36 & 2012 Jan. 19 15:24 & 2012 Jan. 23 07:36 / \\
14:32 -- 21:40 & & & 2012 Jan. 22 17:00 \\
M3 AR1402 & & & \\
\hline
2012 Jan. 23 & 2012 Jan. 23 04:00 & (A)2012 Jan. 23 04:24 & 2012 Jan. 25 15:40 / \\
01:42 -- 07:10 & & (B)2012 Jan. 23 03:24 & 2012 Jan. 24 21:00 \\
X1 AR1402 & & & \\
\hline

\end{tabular}
\end{table}

\subsection{Interaction of CME/ICME with HSS on the basis of interplanetary measurements} 
  \label{Interaction of CME/ICME with HSS on the basis of interplanetary measurements}

Figures 2 and 3 show the interplanetary plasma and field parameters and magnetospheric indices for two sequential Carrington rotations CR 2118 and 2119 (1) 27 December 2011 -- 3 January, 2012 and (2)  25 January -- 1 February 2012. The figures present the following 1-hour parameters (panels from top to down).

First panel: the beta parameter -- the ratio of thermal and magnetic pressures (black line); the T/$T_{exp}$ -- ratio of measured temperature and temperature expected on the basis of average temperature-velocity relation in the solar wind \citep{lopez87} (blue line); the thermal pressure nkT (red line); the ratio of alpha-particle and proton abundances (green line).

Second panel: the magnitude B of IMF (black line); the south-north component $B_{z}$ of IMF (blue line); the radial component $B_{x}$ of IMF (green line); the increment (gradient) of magnetic field magnitude DB on an interval of six hours (red line).

Third panel: the measured proton temperature T (black line); the expected temperature $T_{exp}$ (blue line).

Fourth panel: the density N (black line); the kinetic pressure nV$^2$ (blue line); the increment (gradient) of density DN on an interval of six hours (red line).

Fifth panel: the bulk velocity V  (black line); the increment (gradient) of velocity DV6 on an interval of six hours (blue line).

Sixth panel: the $K_{p}$ index (black line); the component $E_{y}$ of interplanetary electric field (blue line).

Seventh panel: the $D_{st}$ index (black line); the density-corrected $D_{st}$ index \citep{Burton75} (red line); the horizontal red straight line shows the  level of -50 nT.

\begin{figure} 
\centerline{
\includegraphics[width=1.0\textwidth,clip=]{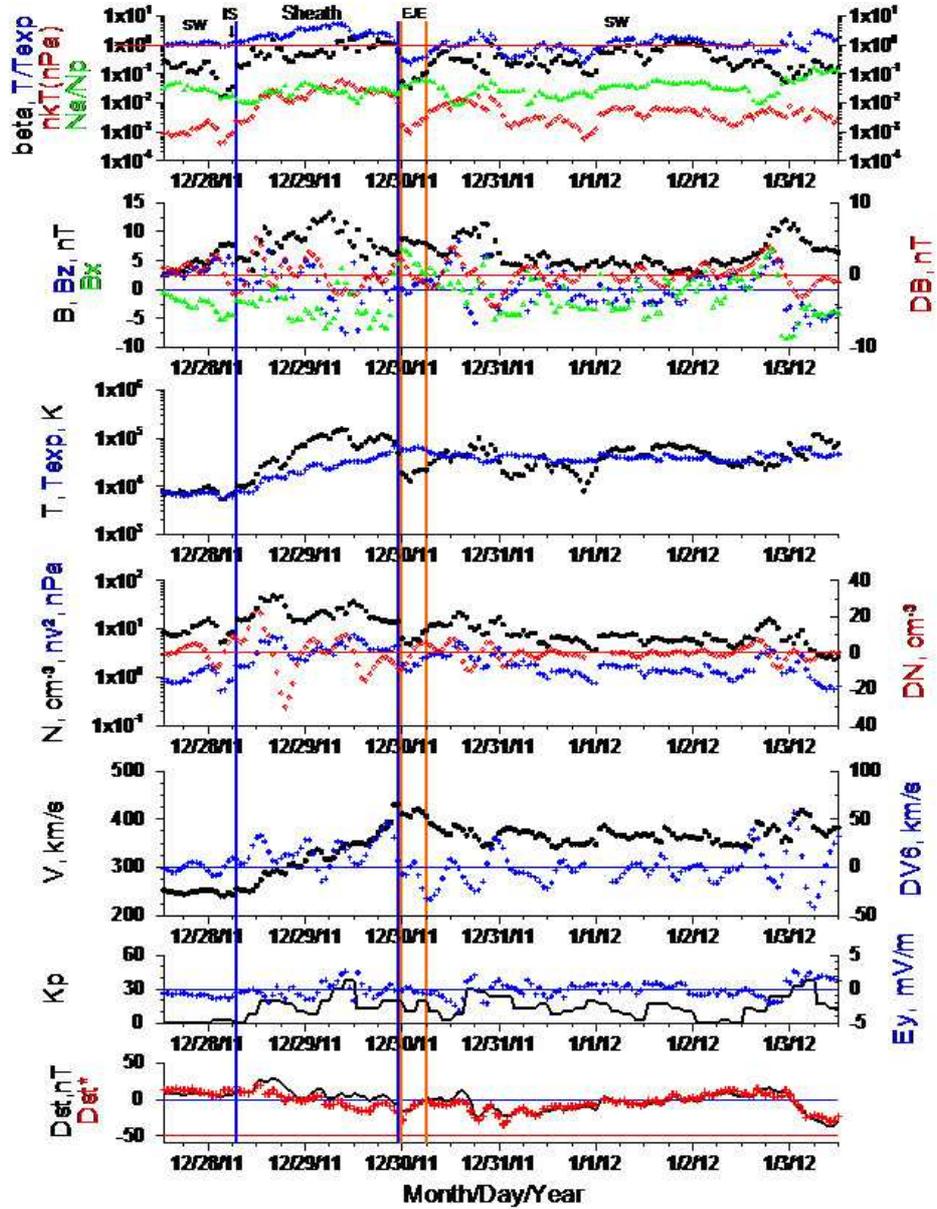}
}
\caption{The interplanetary plasma and magnetic field parameters and magnetospheric indices for interval 27 December, 2011 - 3 January, 2012 (The legend is described in the text). Blue vertical lines show Sheath (the arrow before first blue line indicates the interplanetary shock) and red lines ICME (Ejecta)
}
\label{fig2}
\end{figure}

\begin{figure} 
\centerline{
\includegraphics[width=1.0\textwidth,clip=]{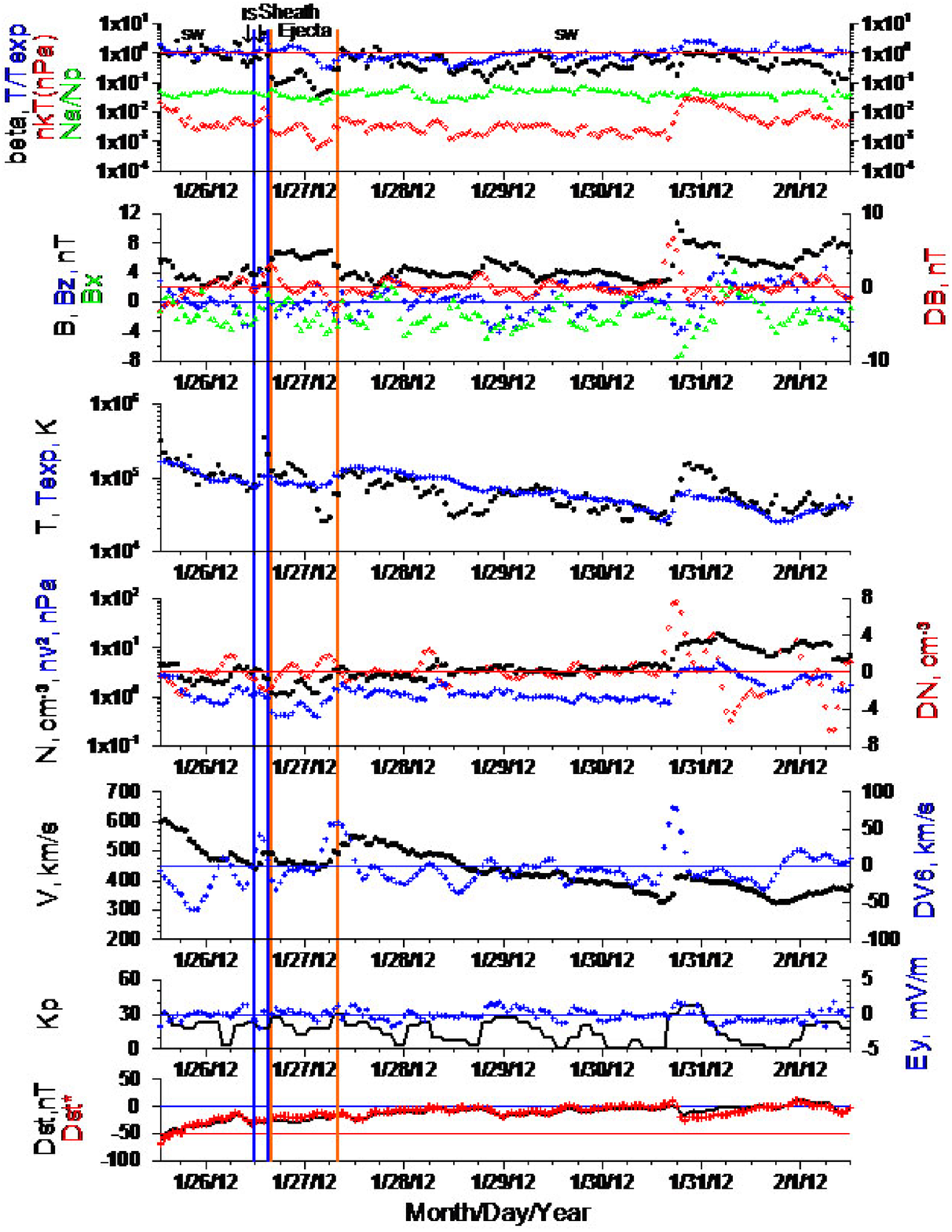}
}
\caption{The same parameters as in Figure 2 for interval 25 January - 1 February, 2012
}
\label{tem_den_evol}
\end{figure}

The identification of solar wind phenomena analyzed in the work is presented on the upper panels in the Figures 2 and 3. In the beginning of figure 2 there is a typical solar wind (indicated as SW). The Sheath is observed on  28 December 2011 at 06:00 UT -- 29 December 2011 at 22:00 UT (indicated by 2 blue vertical lines) and the Ejecta  is registered on 29 December 2011 at 23:00 UT -- 30 December 2011 at 05:00 UT. Just after the Ejecta an atypical solar wind is observed and these data will be discussed latter. Figure 3 demonstrates the same interplanetary structures: undisturbed solar wind, Sheath (26 January 2012 11:00 UT --  26 January 2012 14:00 UT) and  Ejecta (26 January 2012 15:00 UT -- 27 January 2012 07:00 UT). The main difference of Figures 2 and 3 is fact that just after  Ejecta in Figure 3 the HSS begins and there is no HSS in the Figure 2. It is important to note that parameters for 5 hours after first Ejecta can be interpreted as a modification of Ejecta.

If the Ejecta for first rotation 29 -- 30 December 2011 interacted with HSS from the coronal hole and could deflect HSS from Sun-Earth line, the Ejecta would show some consequences of  such interaction. First of all it is necessary to note that the average durations of Sheath and Ejecta during 1976-2000 are about 16$\pm$10 and 30$\pm$20 hours, respectively \citep{yermolaev09} but the durations of first rotation are too long  for Sheath (35 hours) and too short for Ejecta (6 hours) relative to durations relative to mentioned above average durations and very short  duration of Sheath duration  (4 hours) and standard durations of the Ejecta (16 hours) of second rotation. Duration of Ejecta usually correlates with duration of Sheath  and it would be possible to expect that the duration of Ejecta  would be more average size. However the duration of first Ejecta is very short, and this fact indicates that the Ejecta could interact with surrounding streams of plasma.

\begin{figure} 
\centerline{
\includegraphics[width=1.0\textwidth,clip=]{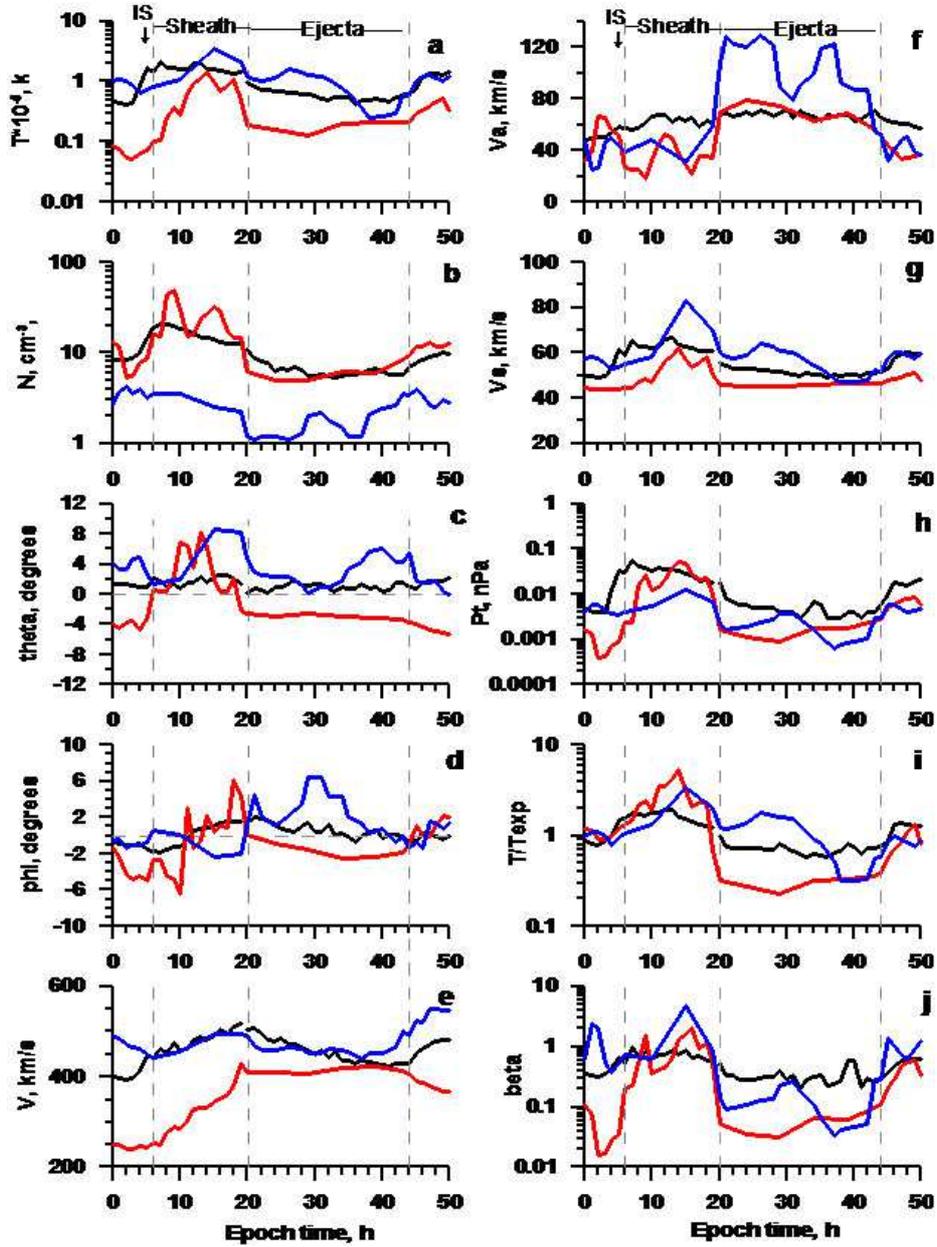}
}
\caption{The temporal profile of interplanetary plasma and IMF parameters and magnetospheric indices for complicated event including interplanetary shock, Sheath and Ejecta obtained by the double superposed epoch analysis (black line), for first solar rotation (red line) and second solar rotation (blue line)
}
\label{tem_den_evol}
\end{figure}

\begin{figure} 
\centerline{
\includegraphics[width=1.0\textwidth,clip=]{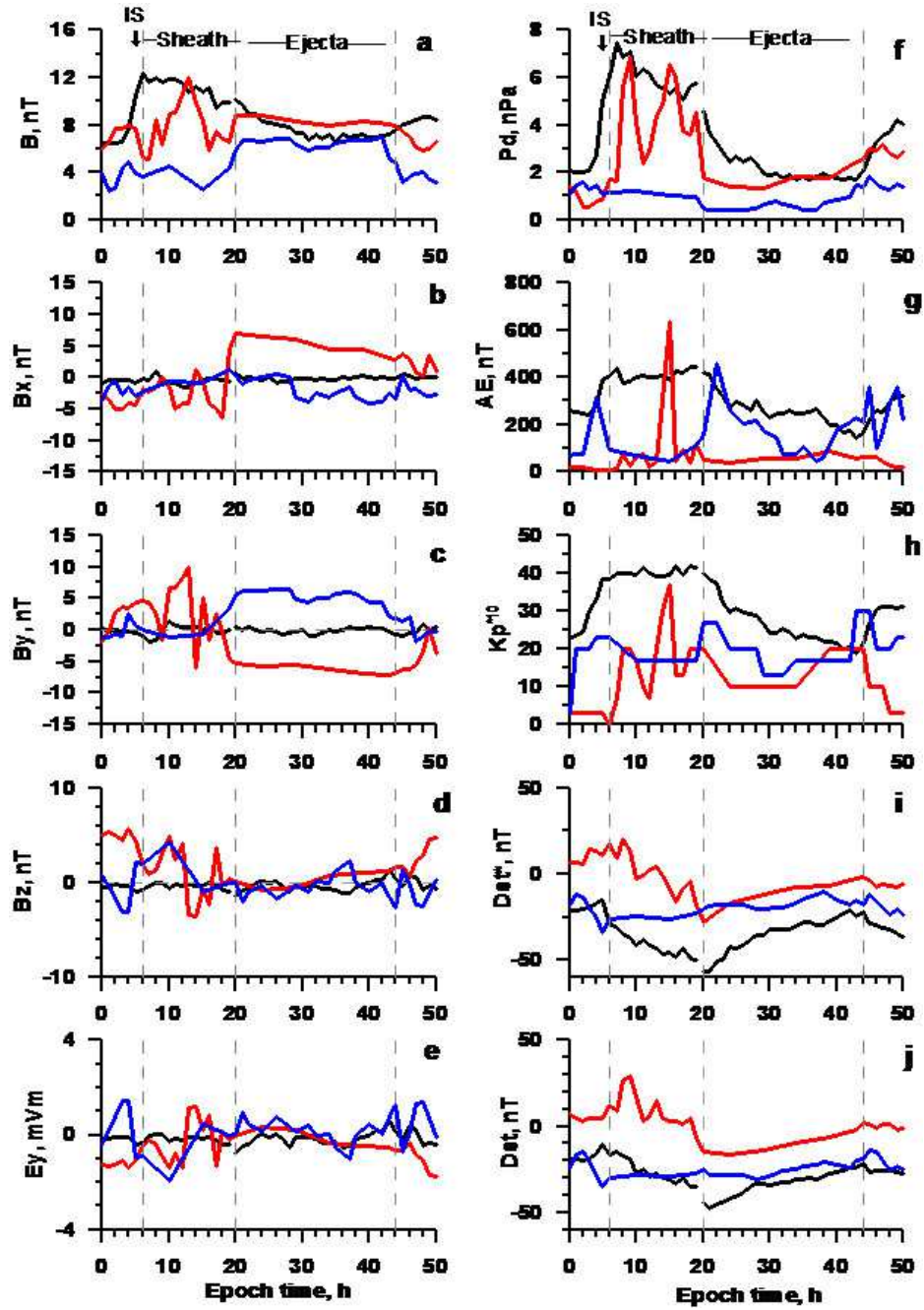}
}
\caption{Continuation of Figure 4
}
\label{tem_den_evol}
\end{figure}

To check the possibility of the interaction we re-scale the durations of Sheaths and Ejecta for both solar rotations and for period 1976-2000 and make equal durations: 14 hours for Sheaths and 24 hours for Ejecta.  Figures 4 and 5 show the temporal profiles of several parameters (destinations are the similar as in Figures 2 and 3) for first interval (red curves), second interval (blue line) and averaged values (black line). The behavior of parameters in Sheath and Ejecta for both intervals is close to average temporal profiles of corresponding parameters in the same phenomena.

It is important to note that parameters during 5-6 hours after first Ejecta (see Figures 2, 4 and 5) can be interpreted as a modification of Ejecta: T, T/$T_{exp}$, N and Pt are higher than average ones in Ejecta and are close in undisturbed solar wind. After this interval (during 30 December 2011 at 11:00 -- 23:00 UT) the compressed solar wind is observed: B, T, T/$T_{exp}$, N and Pt are higher than average ones in undisturbed solar wind. Abrupt change in parameters at 11:00 UT may be an indicator of interface between the Ejecta and solar wind.  As the Figure 4 shows, the latitude speed angle \textit{Theta} is negative during the Ejecta and decreases during the interval after Ejecta.

Such behavior of parameters can be explained with the fact that a back part of Ejecta is pushed, significantly compressed and deflected by faster stream. However, this faster stream was not registered with onboard instruments because of deflection of HSS (or probably a small part of HSS near CME-HSS interface) from the Sun-Earth line.

The relative positions of the sources of HSS and CME streams  (Figure 1, the upper panel) suggests that propagation of the CME streams near the ecliptic plane can deflect the HSS stream to the south by such a way that this stream had not been registered by ACE. Thus, the CH passage through the central meridian with simultaneous occurrence of flares and CMEs in the nearby located sources can result in interaction between HSS and CME streams, deflection of the HSS stream aside of the ecliptic and changes in parameters of the solar wind.

\section{Summary} 
      \label{Summary}
In present work we analyzed the interplanetary parameters near Earth for two consecutive passings of coronal hole in December 2011 -- January 2012 along the Sun-Earth line when these events were preceded by CMEs on the Sun. In both cases the high-speed streams near the Earth have been predicted, however in the first case the high-speed stream near the Earth was not observed, and in the second case it arrived. In the case of absence of HSS, the measured interplanetary parameters show that preceding  interplanetary CME (Ejecta)  was very short and compressed by the subsequent fast  stream and this fact may be considered as an indirect evidence  that the HSS from coronal hole collides with CME and the collision results in compression of Ejecta and deflection of HSS from the Sun-Earth line.

\begin{acks}
 The authors are grateful for the opportunity to use data of the OMNI database (http://omniweb.gsfc.nasa.gov), the Solar Terrestrial Relations Observatory STEREO--A and B (http://helio.gmu.edu/seeds/secchi.php), the Large Angle and Spectrotrometric Coronagraph Lasco (http://cdaw.gsfc.nasa.gov), the Solar Dynamic Observatory SDO/AIA telescope (http://sdo.gsfc.nasa.gov), the SolarDemon database (http:// solardemon.oma.be/ science/flares.php) and the Advanced Drag Model and the WSA--Enlil cone model (http:// helioweather.net). This work was supported by the Russian Scientific Foundation, project 16--12--10062.
\end{acks}




\bibliographystyle{spr-mp-sola}
\bibliography{sola_bibliography_Yermolaev}

\end{article}

\end{document}